# Linear increase of the $2^+$ ion concentration in the double photoionization of aromatic molecules

D. L. Huber

*Department of Physics, University of Wisconsin-Madison, Madison, WI 53706, USA*

We investigate the linear behavior in the $2^+$ ion concentration observed in the double photoionization of a variety of aromatic molecules. We show it arises when the photoelectrons are emitted simultaneously. Neglecting the momentum of the incoming photon and the momentum transferred to the molecule, it follows that the momenta of the individual photoelectrons are oppositely directed and equal in magnitude. Under steady-state conditions, the ion concentration is proportional to the rate at which the ions are created which, in turn, varies as the product of the densities of states of the individual electrons. The latter vary as the square root of the kinetic energy, leading to overall linear behavior. The origin of the linear behavior in pyrrole and related molecules is attributed to the presence of atoms that destroy the periodicity of a hypothetical carbon loop. In contrast, the resonant behavior observed in pyridine and related molecules, where a fraction of the CH pairs is replaced by N atoms, is associated with electron transfer between the nitrogen atoms and carbon atoms that preserves the periodicity of the closed loop.

**I. Introduction**

In a series of recent papers[1-5], R. Wehlitz and his colleagues presented results for the photon energy dependence of the $2^+$ ion concentration associated with the double photoionization (DPI) of a variety of aromatic organic molecules. Broadly speaking, the results fall into two categories. With the knockout contribution removed, the cyclic hydrocarbons showed a resonant peak in the $2^+$ ion concentration at the onset of the DPI, whereas molecules where one or more of the carbon atoms were replaced by non-carbonic atoms frequently displayed a linear increase in the $2^+$ concentration with photon energy.[2] Recent studies of pyrrole and benzene by K. Jänkälä *et al.* involving electron coincidence measurements revealed that in pyrrole, a molecule showing linear behavior in the $2^+$ concentration at the onset of DPI, (Fig. 1), the two photoelectrons are emitted simultaneously.[6]

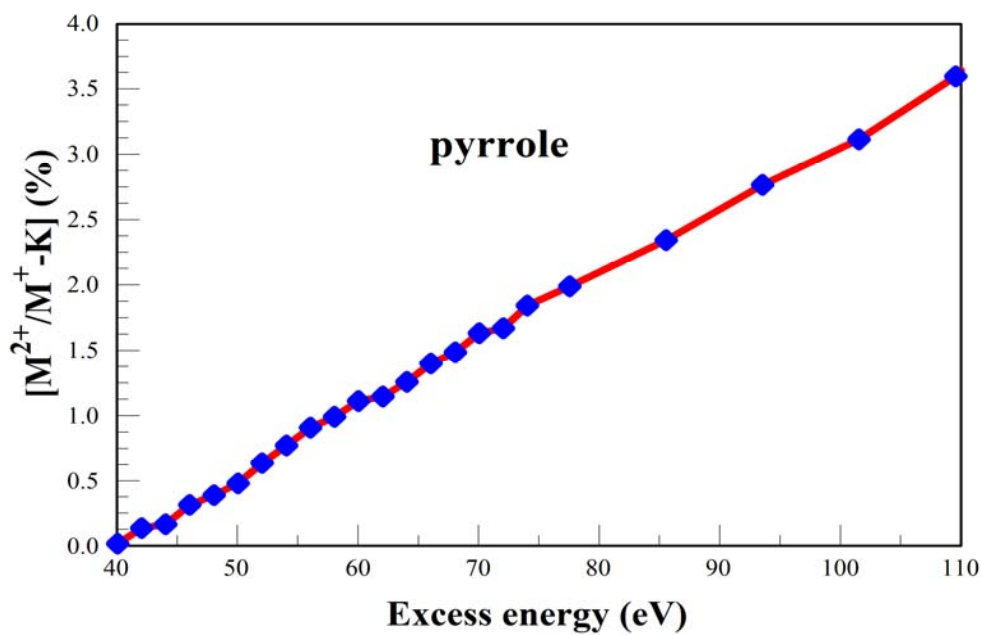

Fig. 1. Pyrrole. $2^+$ ion ratio, with knockout contribution (*K*) subtracted, vs excess energy (photon energy above knockout threshold).[2] Data provide by R. Wehlitz.

## II. General Analysis

The starting point in the analysis of the linear behavior is the connection between the rate at which the 2$^+$ ions are created in the DPI, which we designate by $R$, and their concentration, $M^{2+}$. Under steady-state conditions, the rate and the concentration are related by the equation

$$R = (1/\tau^{2+})M^{2+} \tag{1}$$

where $\tau^{2+}$ denotes the average lifetime of the 2$^+$ ions. In the case of single-electron ionization, the rate at which the ions are created is proportional to the density of energy states of the emitted electron which, in turn, is proportional to the square root of the kinetic energy of the electron.[7] In the case of double photoionization where the electrons are emitted simultaneously, the production rate is proportional to the product of the densities of states of the two particles.[8] Thus from Eq. (1) we have

$$M^{2+} \propto (E_1 E_2)^{1/2} \tag{2}$$

where $E_1$ and $E_2$ are the kinetic energies of the two electrons.

By making use of the conservation of energy and momentum, one can show that $E_1 = E_2$. The momentum of the photon involved in the excitation is negligible, as is the momentum transferred to the ion.[9] Neglecting both the photon and ion momenta, the momentum conservation equation for the electrons reduces to

$$0 = \mathbf{p_1} + \mathbf{p_2} \tag{3}$$

leading directly to back-to-back emission with equal kinetic energies. As a result, the 2$^+$ concentration depends linearly on the photon energy, as shown in Fig. 1. It should be noted that this analysis applies to all double photoionization transitions where the two photoelectrons are emitted simultaneously with zero total momentum not just organic aromatic hydrocarbons. For example, a recent publication by Wehlitz et al. reported linear behavior in the double photoionization of the inorganic molecule tribromoborazine.[10]

## III. Discussion

The authors of Refs. 1-5 attributed the resonant peaks in the DPI to the formation of a mobile, two-electron quasi-bound state. In their interpretation, the photoionization is a two-step process in which a quasi-bound state is created and subsequently decays into two free electrons. The question then arises as to the nature of the quasi-bound state. In our earlier paper[8], we proposed a mechanism that involved the formation of Coulomb pairs at the DPI threshold where 'Coulomb pair' refers to the quasi-bound state of two electrons in a one-dimensional array with periodic boundary conditions.[11] In the present context, the one-dimensional periodic array is associated with the carbon atoms on the perimeter of the molecule, and the threshold state is associated with a carbon site occupied by two π electrons with antiparallel spins

The molecules pyrrole, furan and selenophene are comprised of a five-site ring, where four of the sites are occupied by CH pairs and the fifth by an 'impurity'. The impurities are a NH pair (pyrrole), an O atom (furan) and a Se atom (selenophene).[5] Equivalent results have been reported recently for tropone where a CH site in a seven-member ring is replaced by an O atom.[12] In all four cases, the threshold behavior is marked by linear increases in the $2^+$ ion concentration. We attribute the linear behavior to the impurity which introduces a gap in the array blocking excitation transfer between carbon atoms on opposite sides of the impurity. The effect of this gap is to eliminate the pairing by converting a hypothetical five-carbon closed loop to a four-carbon, or in the case of tropone a six-carbon, chain plus an impurity site that blocks the formation of the Coulomb pair.

In the molecules mentioned in the preceding paragraph, the replacement of a CH pair by an impurity eliminates Coulomb pairing. However, in a recent publication, Hartman and Wehlitz showed that when CH pairs in benzene are replaced by nitrogen atoms, Coulomb pairing is retained.[13] The molecule 1, 3, 5-triazine has three nitrogen atoms that alternate with three CH pairs, whereas pyridine has one nitrogen atom, and pyrimidine, two nitrogen atoms separated by a CH pair. The peak in 1, 3, 5-triazine is approximately the same size as the peak in benzene but shifted to higher energy by about 10 eV. In the case of pyridine, the peak is much smaller but is at the same location as the peak in 1, 3, 5-triazine (Fig. 2). From these studies, we draw the conclusion that excitation transfer from CH pairs to nitrogen in the 6-site ring is allowed thus making possible Coulomb pairing in all three molecules.

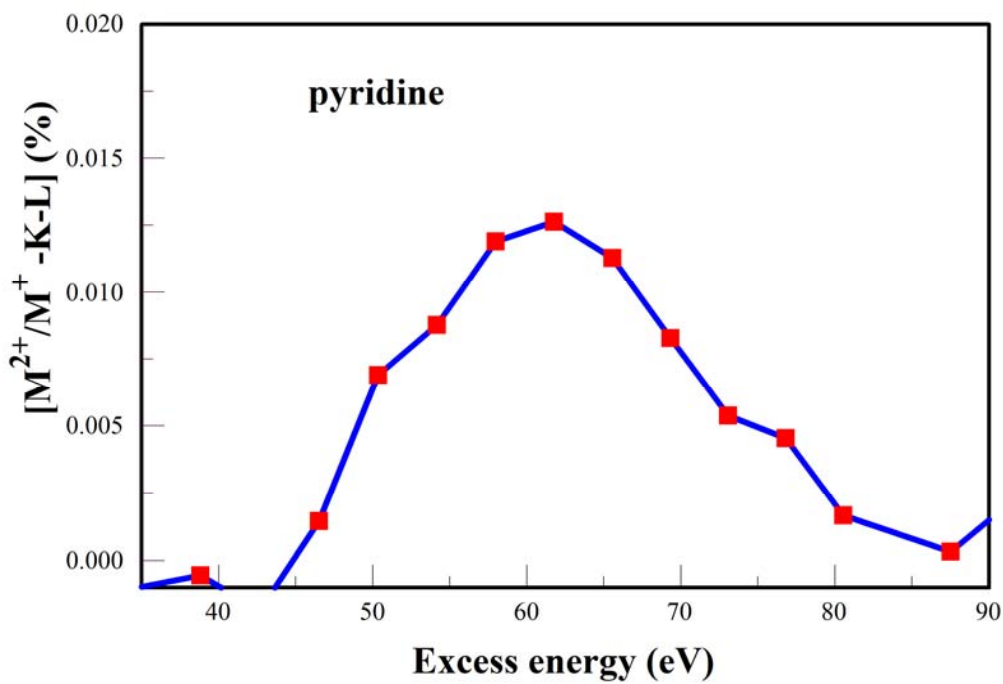

Fig. 2. Pyridine. 2$^+$ ion ratio, with knockout (*K*) and quasi-linear fit (*L*) contributions subtracted, vs excess energy (photon energy above knockout threshold). The quasi-linear fit contributions are associated with the emission of electrons with approximately equal kinetic energies. Data provide by R. Wehlitz.

**Acknowledgment**

We would like to thank Ralf Wehlitz for helpful comments and for providing the data displayed in Figs. 1 and 2.

**References**

1. R. Wehlitz, P. N. Juranic, K. Collins, B Reilly, E. Makoutz, T. Hartman, N. Appathurai, and S. B. Whitfield, *Phys. Rev. Lett.* **109** (2012) 193001.
2. T. Hartman, P. N. Juranic, K. Collins, B Reilly, E. Makoutz, N. Appathurai, and R. Wehlitz, *Phys. Rev. A* **87** (2013) 063403.
3. T. Hartman, K. Collins, and R Wehlitz, *Phys. Rev. A* **88** (2013) 024701.
4. R. Wehlitz and T. Hartman, *J. Phys. Conf. Series* **488** (2014) 012013.
5. R. Wehlitz, *J. Phys. B: At. Mol. Opt. Phys.* **49** (2016) 222004.
6. K. Jänkälä, P. Lablanquie, F. Penent, J. Palaudoux, L. Andric, and M. Huttula, *Phys. Rev. Lett.* **112** (2014) 143005.
7. H. A. Bethe and R. W. Jackiw, *Intermediate Quantum Mechanics, Third Edition*; (Benjamin/Cummings: Menlo Park, 1986).
8. D. L. Huber, *Phys. Rev. A* **89** (2014) 051403 (R).
9. M. Ya. Amusia, E. G. Drukarev, V. G. Gorshkov, and M. P. Kazachkov, *J. Phys. B* **8** (1975) 1248.
10. R. Wehlitz, M. MacDonald, L. Zuin, A. C. F. Santos, and N. Appathurai, *Phys. Rev. A* **95** (2017) 023408.
11. S. M. Mahajan and A. Thyagaraja, *J. Phys. A: Math. Gen.* **39** (2006) L667.
12. T. Hartman and R. Wehlitz, *J. Chem. Physics* **146** (2017) 204306.
13. T. Hartman and R. Wehlitz, *Phys. Rev. A* **91** (2015) 063419.